# FindPOTATOs: Minor Planet Observation Linking Software

C. R. Nugent[1] , Nicole J. Tan[2] , and James M. Bauer[3]
[1] Olin College of Engineering, 1000 Olin Way, Needham, MA 02492, USA; cnugent@olin.edu
[2] School of Physical and Chemical Sciences—Te Kura Matū, University of Canterbury, Private Bag 4800, Christchurch 8140, New Zealand
[3] Department of Astronomy, University of Maryland College Park, MD 20742, USA

## Abstract

For minor planet observations to be archived and used by the scientific community, observations of individual objects must be linked together into groups called tracklets. This linking is nontrivial, as linking software must find real tracklets in noisy data within a reasonable amount of time. We describe FindPOTATOs, a linking software written in Python that can assemble minor planet tracklets. With the appropriate parameters, FindPOTATOs assembles tracklets for a variety of objects, including close-approaching near-Earth objects and trans-Neptunian objects. FindPOTATOs is ideal for processing data sets taken at small observatories, processing archival data, or finding tracklets of minor planets on unusual orbits. This paper describes the code structure, usage, and validation.

## 1. Introduction

Minor planets are generally visually indistinguishable from other astronomical objects, such as stars. Therefore, minor planets have traditionally been discovered based on their movement across the sky relative to background stars. This was the case for Giuseppe Piazzi, who discovered Ceres by noticing a moving object in a star map he drew over several nights (G. Foderà Serio et al. 2002), and it is still the case for today's asteroid-discovery surveys.

The Minor Planet Center (MPC),[4] based in Cambridge, MA, USA, is the worldwide archive and arbiter of minor planet observations. In general, observations must be submitted to and accepted by the MPC to be useful to the broader scientific community. In most cases, the MPC will not accept single observations of minor planets.[5] Observations are usually accepted in the form of a set of observations of a single object. This set of observations is called a "tracklet." Tracklets must be comprised of at least two observations; at least three or four observations are preferred to prevent false linkages.

Accordingly, observers must assemble tracklets from their data if they would like to submit their observations to the MPC. Tracklet assembly can be done manually or with the assistance of linking software.

Linking software was first developed alongside the adoption of digital sensors in astronomical surveys. One of the first digital sky surveys, Near Earth Asteroid Tracking (NEAT), created a series of subroutines in FORTRAN and C to find asteroids (S. H. Pravdo et al. 1999). Candidate sources were assembled via a series of routines, some of which would reject stationary sources (TABMATCH) and perform further cleaning to remove clusters of detections associated with diffraction spikes (TABEDIT). These detections were then assembled into candidate tracklets by NEOFIND. Unfortunately, we have not been able to find an archived version of this software.

The Moving Object Pipeline System (MOPS; J. Kubica et al. 2007; L. Denneau et al. 2013) is a popular algorithm used by Pan-STARRS that associates detections with approximate linear minor planet sky motions. This algorithm has been adapted by several active surveys including NEOWISE (R. M. Cutri et al. 2013) and Asteroid Terrestrial-impact Last Alert System (ATLAS; J. L. Tonry et al. 2018). ATLAS developed and now employs the algorithm PUMALINK, which can evaluate 10 million possible tracklets in a half hour of computer wall-clock time (J. L. Tonry 2023). M. J. Holman et al. (2018) developed an algorithm called HelioLinC. HelioLinC performs coordinate transformations, assuming that the minor planets of interest are at roughly a given distance from the Sun. These transforms align the detections in the new coordinate space, simplifying linking. This algorithm has been implemented in the HelioLinC3D code base, which will be part of the Vera C. Rubin Observatory Legacy Survey of Space and Time survey pipeline and has been tested against other data sets (e.g., A. Heinze et al. 2023). Another algorithm, THOR (J. Moeyens et al. 2021), uses coordinate transforms and precomputed orbits to link detections. Optimized for main-belt objects, it has not been adapted to near-Earth objects (NEOs). Finally, CANFind was developed to identify spatial groups of observations within the NOIRLab Source Catalog (K. M. Fasbender and D. L. Nidever 2021). Historically, the code implementation of linking algorithms used by asteroid surveys were not made public; this recently has changed with the release of the PUMA, HelioLinC3D, THOR, and CANFind code on GitHub.

The software implementations of MOPS, PUMALINK, and HelioLinC3D are all tuned for speed and accuracy for use in major surveys. Asteroid surveys operating nightly must submit their observations to the MPC in a timely manner so that new discoveries can be followed up by other observatories. Therefore, their linking software must handle huge amounts of data quickly and accurately. For example, ATLAS surveys the visible sky four times each night, corresponding to over 100,000 $\deg^2$/day (J. L. Tonry 2023).

---

[4] https://www.minorplanetcenter.net/
[5] There are rare exceptions to this. Single observations have been accepted; for example, when the detection can be shown to significantly improve an orbit and there is evidence that the detection is real and not spurious. However, the bulk of MPC-submitted observations are in tracklet format.

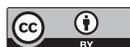

**Table 1**
List of Columns That are Required and Optional for FindPOTATOS Source Input

| Column Name | Value | Required? |
| --- | --- | --- |
| RA | Float. R.A. of detection, degrees. | Yes |
| Dec | Float. decl. of detection, degrees. | Yes |
| magnitude | Float. Brightness of detection, in magnitudes. | Yes |
| mjd | Float. Date and time of midpoint of observation, in Modified Julian Date format. This value should be the same for all detections in a single frame. | Yes |
| observatory_code | String. MPC-assigned observatory code where observations were taken. | Yes |
| band | String. MPC-defined observing band (wavelength) of detections | Yes |
| mag_err | Float. Uncertainty on magnitude, as defined by ADES export format. | When using ADES export format. |
| RA_err | Float. Uncertainty on R.A., as defined by ADES export format. | When using ADES export format. |
| Dec_err | Float. Uncertainty on decl., as defined by ADES export format. | When using ADES export format. |
| ml_probs | Float. The probability that the detection is true; assigned by a machine learning algorithm. | No |

The currently available linking software is highly tested, efficient, and by necessity specialized. Missing from this software ecosystem is an open-source linking software that is simple to use and easily adapted to a range of data types and target minor planets.

We describe Find Point-source Object Tracklets Affirmed Through Orbit-fits (FindPOTATOS). The software is available on GitHub;[6] a frozen version archived at the time this paper was written is also available (C. R. Nugent 2024). When provided the appropriate parameters, FindPOTATOS can identify tracklets for a range of minor planet types, from close-approaching near-Earth objects to Trans-Neptunian objects. Possible use cases include:

1. *Examining archival data.* Planetary defense and solar system science in general benefit from the culling and improvements of detections from archived astronomical data. Images that are archived and publicly accessible can more than double the number of science papers produced from a data set (R. L. White et al. 2009). Identified minor planet detections provide a wealth of data that are more regularly sampled and can be debiased with the purpose of extrapolating sampled populations and population characteristics. Most archived data have not been searched for minor planets; reporting asteroid detections from archival data can provide valuable precovery data and secure orbits of potentially hazardous objects.

2. *Identifying tracklets of minor planets on edge-case orbits.* Automated survey linking software used by discovery surveys is generally tuned to specific classes of orbits, such as near-Earth objects. Software engineers and astronomers are careful to ensure high recall and accuracy of the linking software, and the resulting tracklets generally include other classes of orbits (such as main-belt objects) as well. However, there are cases where a researcher may wish to examine a data set to find a particular class of objects that were not previously detected. For example, HelioLinC3D gains efficiency and speed from frame transformations and is most effective in detecting objects that do not have significant Sun–object distance changes over the period of observation; a researcher interested in high eccentricity comets may choose to use FindPOTATOS, with the appropriate parameters, to find those objects. Similarly, although PUMALINK is highly efficient at searching ATLAS data for NEOs, it is not targeted toward distant objects such as trans-Neptunian objects (TNOs). A researcher may choose to search a subset of ATLAS data, carefully selected at an appropriate cadence, and use FindPOTATOs to link TNO observations.

3. *Reporting minor planet observations in a small data set.* Observers may obtain a series of images of the sky for another scientific or recreational purpose and want to find minor planets in the data set. Students at school observatories may wish to observe minor planets and contribute their data to the MPC, so that it may be used by the broader scientific community. In the spirit of open science, lowering the barriers to these groups creates powerful educational experiences and unlocks more complete usage and possibly unique applications of those data sets.

FindPOTATOs was written in Python for ease of use and ease of adaptation by users and is available via the MIT License. It accepts input data in comma-separated values (CSV) format. CSV format was chosen for its simplicity and compatibility with most Python libraries.

## 2. Formats for Data Input and Export

The MPC accepts tracklets in two formats: 80-column and Astronomy Data Exchange Standard (ADES; ADES 2024). The 80-column format is a historical standard that is useful for some tools (e.g., the MPC's Minor Planet Checker[7]); ADES is preferred for submissions, as it includes measurement uncertainties as well as other additional information. FindPOTATOS can export tracklets in both formats.

Asteroid discovery surveys generally observe the same region of sky multiple times over a single night. The data is reduced, bright points, or sources, are detected in the image, stationary sources are removed, and then the resulting sources are processed via a linking algorithm to create tracklets.

FindPOTATOS is designed to handle CSV files of detected sources; we refer to these individual files as "frames," as for most users these are the sources corresponding with one frame

---

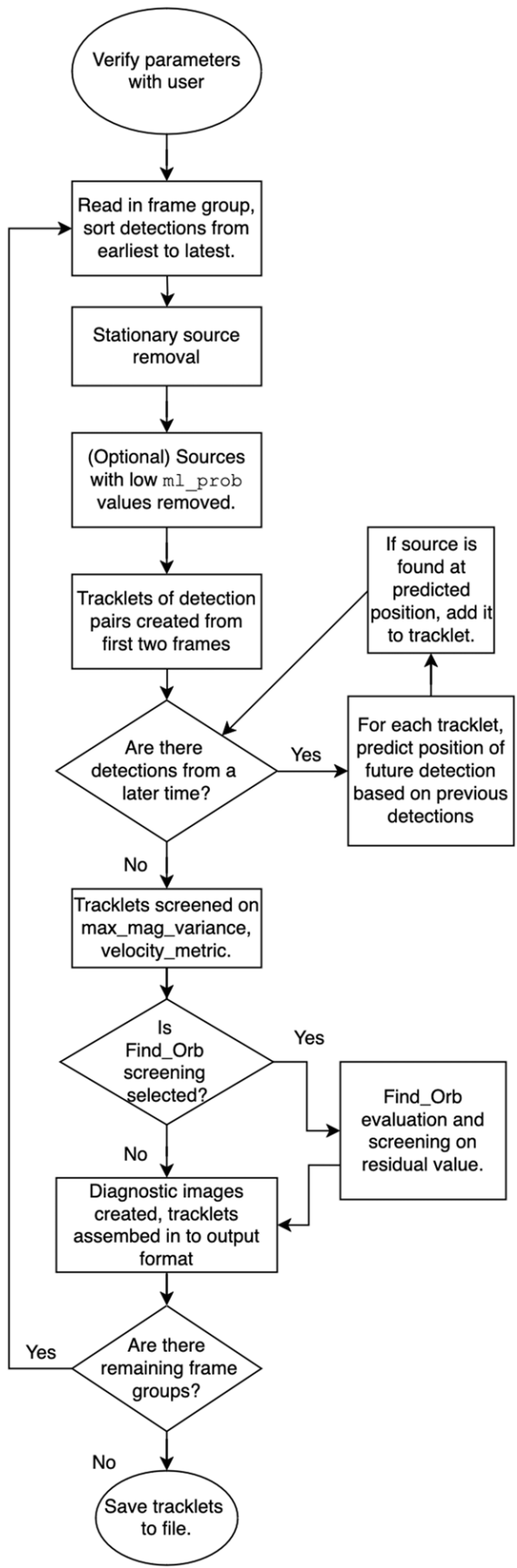

**Figure 1.** Schematic of FindPOTATOs.

image of the sky. Each frame contains sources from the same region of sky, taken at a single time.

As input, FindPOTATOs requires another CSV file, which lists the frames that cover the same region of sky. We refer to these as "frame groups." For most users a frame group would be the set of frames taken using the same telescope pointing. Tracklets are assembled from within the frames in a frame group; FindPOTATOs will not link detections between frame groups. Each list of frame groups should be in the format of comma-separated rows, where each row represents one region of sky, and each column is the name of the frame file.

Each row in the frame CSV file contains information about a single detection with columns described in Table 1. To create such a file, sources can be extracted from images using standard algorithms such as DAOPHOT (P. B. Stetson 1987) or SExtractor (E. Bertin & S. Arnouts 1996); Python implementations of these algorithms (Astropy Collaboration et al. 2013, 2018, 2022; K. Barbary 2016) are available. It is vital that stationary sources (such as stars) is removed from the data; this can be accomplished via image differencing or comparison of detected sources to a known star catalog.

The transient sources remaining in a data set after star removal often contain sources that are not minor planets; this can include glints from bright stars, cosmic rays, galaxies and satellites. To further clean the data, many discovery surveys employ screening using machine learning or neural net algorithms; these trained classifiers can assign a probability that a source of interest has the physical morphology of a minor planet. FindPOTATOs can screen sources when given a threshold probability of correctness, if each input source has an associated probability provided in the ml_probs column. This column could be used to screen on other metrics, such as signal to noise of the detection.

Table 1 lists the columns that can be present in the CSV file. If additional columns are present, they will be ignored by the code. The columns do not need to be in order. Uncertainties on measurements are not used by the algorithm, but they are reported in the ADES export format, if that export format is chosen.

The ADES format allows for inclusion of additional information about contributors to the measurements, the telescope used, funding source, and so forth. These must be updated by the user in parameters.py, following the ADES guidelines.

## 3. Code Description

FindPOTATOs' linking algorithm is based on previous work, including NEOFIND (S. H. Pravdo et al. 1999) and MOPS (L. Denneau et al. 2013). An overview of the code is shown in Figure 1.

FindPOTATOs is designed to process a full night of observations from a sky survey at once. As input, FindPOTATOs requires a list of frame group source files, as described in the previous section.

The code first performs stationary source removal. If any two sources are within the distance specified via parameter stationary_dist_deg, they are removed. While this removes sources that are in the same location in the sky, some stationary sources can be missed. For example, a star that was too faint to be detected in frames A and B but was detected in frame C due to better sky conditions when that frame was taken will not be removed by this stationary source removal.

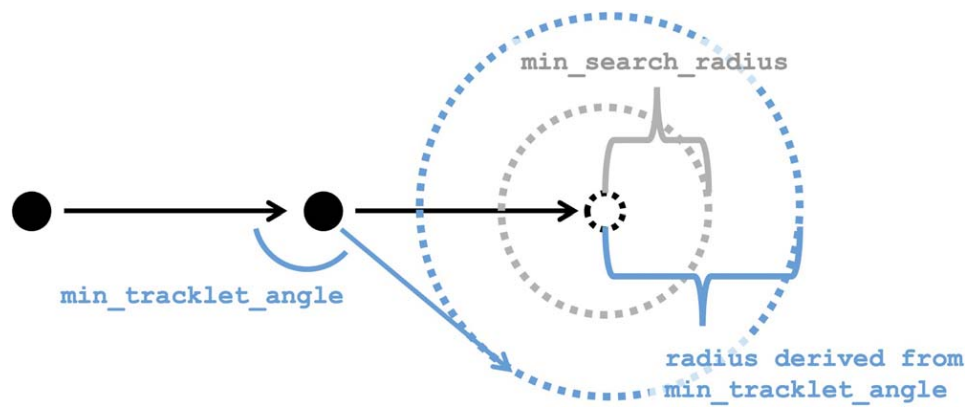

**Figure 2.** The previous two detections in a tracklet (solid black circles) and assumed linear motion are used to predict the position of the next detection in the tracklet (black dotted circle). The code searches around the predicted position, using one of two calculated radii. A minimum search radius (gray dotted circle), min_search_radius, is a parameter set by the user, and is always searched. A secondary radius is also derived from the min_tracklet_angle. This parameter allows tracklets to curve. A curving tracklet may fall outside the min_search_radius and require a larger search radius (blue dotted circle) is then used to find subsequent detections.

Next, all possible pairs of detections between sources in the first and second frames are compiled into a dictionary of Pandas Dataframes (W. McKinney 2010). A possible pair is defined as two sources where the distance between the sources is less than a threshold established by the time between exposures and the parameter max_speed, which is the maximum speed of objects of interest given in units of arcseconds/second.

Detection pairs serve as a vector that allows the position of a possible detection in the next frame to be predicted; this prediction assumes linear motion. FindPOTATOs determines a center position to search for a possible linkage detection, as well as a search radius (Figure 2). The search radius is never smaller than the user-set parameter min_search_radius. If the user seeks tracklets that are not linear; they may choose to set the parameter min_tracklet_angle to a value such as 160°. A curving tracklet may require a search radius larger than min_search_radius to find the next detection; the necessary larger radius to search is calculated by and employed by the code.

The above steps are accelerated via the use of a Ball Tree data structure (T. Liu et al. 2006), implemented using the Scikit-learn Python library (F. Pedregosa et al. 2011). This data structure organizes the sources in each file for efficient retrieval; it was implemented over other structures, such as KD trees, because it was both efficient and easily modified to use a custom distance function to evaluate the distance between two points. The distance function used is based on the Astropy separation() function, which allows accurate angular separation measurements, even near the poles.

If a detection is found within the search radius around the predicted location, it is added to the tracklet. If multiple detections are found in the search radius around the predicted location, then multiple tracklets are assembled.

This process repeats for each of the frames in a frame group. There is no upper limit on the number of detections that can be linked. It is possible that computer memory issues may prevent linking of millions of tracklets, however, that limit is many orders of magnitude above the linking goals of most users today.

Because the code uses the previous two detections to predict a future detection, there must be a continuous chain of detections throughout the frames. Consider a set of five frames, A, B, C, D, and E. FindPOTATOs will produce tracklets that span A–B–C, A–B–C–D, and A–B–C–D–E. It will not produce linkages between B–C–D or B–C–E. To find a linkage of B–C–E, a user could re-run FindPOTATOs with only B–C–E frames.

Users may wish to screen tracklets using various parameters. FindPOTATOs has built in screening around the following categories.

### 3.1. Magnitude Limits

Some minor planets are not expected to change brightness significantly between observations; FindPOTATOs can automatically reject tracklets with brightness variations greater than a specified max_mag_variance value. For example, a limit of 1.5 mag change in brightness can be useful for linking detections of main-belt asteroids over the span of one hour. However, this screening should not be used for objects with expected large changes in brightness, such as very close NEOs.

### 3.2. Velocity Metric

Some minor planets are not expected to change their sky-plane angular velocity significantly over the course of a tracklet. For these objects, FindPOTATOs can screen on velocity_metric, which is the standard deviation of the velocity between each linked detection pair. If set too low, this metric can exclude close-Earth approachers, which can display plane-of-sky apparent velocity changes between detections.

### 3.3. Orbital Fit Screening

Optionally, the tracklets can also be screened via Find_Orb,[8] if that software is installed on the computer running FindPOTATOs. Find_Orb can fit Keplerian orbits to observations. A tracklet is provided to Find_Orb and the residual from an orbit fit is calculated. The user can specify a residual threshold (maximum_residual); below that threshold the tracklet will be kept, above that threshold it will be rejected. Screening via Find_Orb adds minimal additional time. An analysis shows that Find_Orb can screen a candidate tracklet in 0.02 s; for a full night of survey data with 3000 tracklets, this would add an additional minute of running time.

---

[8] https://www.projectpluto.com/find_orb.htm

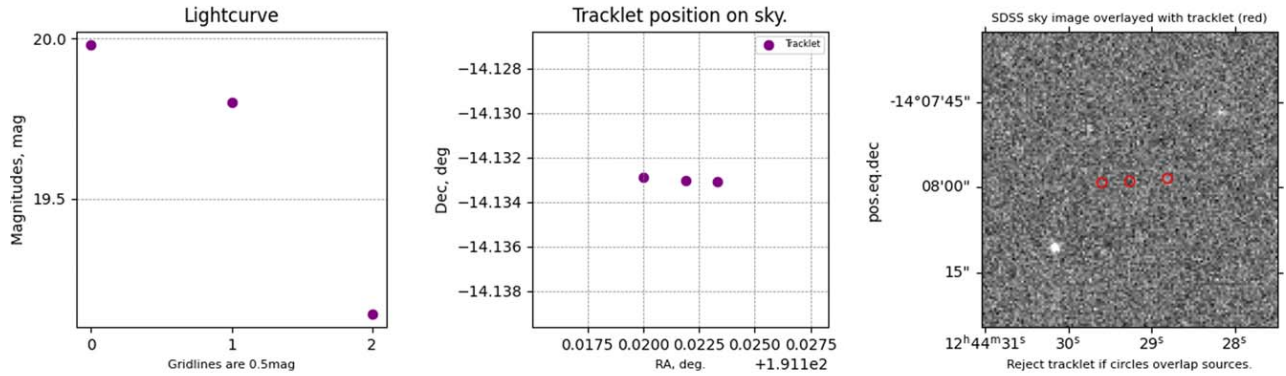

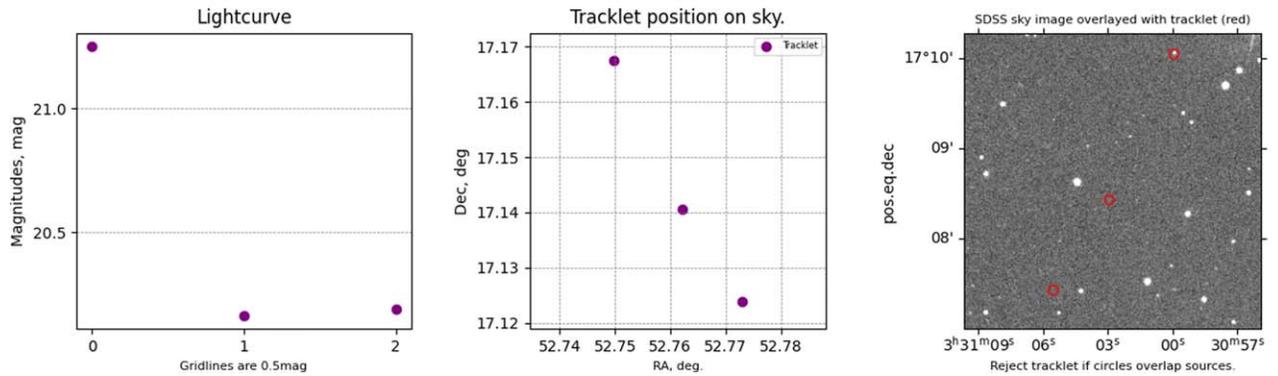

**Figure 3.** Sample diagnostic test images. Top: a real tracklet, from an example data set provided to users based on Catalina Sky Survey data from the MPC. For the three detections in this tracklet, the diagnostic image shows their magnitudes (left), their positions on the sky (middle), and an SDSS image of the same region of sky (right). This last frame allows users to visually verify that the tracklet does not overlap with stationary sources, such as stars. Bottom: a spurious tracklet, where the SDSS image indicates that one of the detections is not a moving object, but a star. This tracklet should not be submitted to the MPC and indicates that the data should undergo further cleaning.

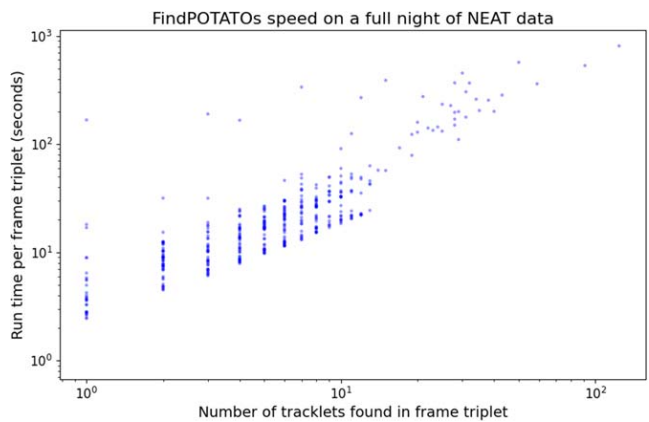

**Figure 4.** NEAT surveyed the sky using a cadence of three images, or frame triplets. The average number of sources in each frame was 728 but varied widely from frame to frame, due to the unique characteristics of the TriCam cameras. The frame with the least sources contained 78 sources, the frame with the most contained 13,319. Therefore, there was significant runtime variation between frame triplets. The runtime per frame triplet generally increased with the number of tracklets found. Frame triplets with zero tracklets found can take tens of seconds to process.

The advantage of integrating Find_Orb with a linking program is that FindPOTATOs can find linkages while being agnostic to orbit type. If the parameters are not artificially limiting (such as max_mag_variance = 0.2), FindPOTATOs can find objects on a range of orbits simultaneously.

### 3.4. Features for Assessing Tracklet Quality

The tracklets are output in the two MPC-approved formats, 80-character and ADES. Diagnostic tracklet images are optionally generated (see Figure 3), showing the tracklet magnitude variations, the path of the tracklet across the plane of the sky, and, if requested by the user, the tracklet position relative to an image of the stationary sky. The default sky image is from the Sloan Digital Sky Survey (SDSS; J. Kollmeier et al. 2019) retrieved via AstroQuery (A. Ginsburg et al. 2019). SDSS does not cover the entire sky; if an SDSS image does not exist, the sky image will display as a black square. If desired, users can modify the call to function query_skyview() to access any other survey available through AstroQuery's SkyView function.

Code speed varies depending on the parameters chosen. An analysis of the running time indicates that the SDSS sky image query and diagnostic image generation is the longest optional parameter. This option takes 0.55 s per tracklet, for a full night of survey data that contains 3000 tracklets this would add another half hour to the running time.

The MPC will reject a tracklet submission if multiple tracklets contain the same observation. As one observation cannot be part of two separate object tracklets, this ensures data quality. After FindPOTATOs is complete, the resulting ADES file can be checked for duplicates using check_repeats.py. This routine displays the diagnostic test images for tracklets that share a detection and allows the user to select which tracklet to keep or allows the user to discard all tracklets. A new ADES file with unique tracklets corresponding to the user's selection (as well as the tracklets without duplicate observations) is produced.

## 4. Code Validation and Testing

FindPOTATOs has been validated against a variety of data sets. Most extensive has been its use in reprocessing the NEAT TriCam data set, as part of a larger reprocessing effort. NEAT was one of the first surveys to use digital sensors for asteroid discovery (S. H. Pravdo et al. 1999). Therefore, this data is noisy, with more digital artifacts than are present in data from modern optical sensors. There can be thousands of sources per frame triplet, with the vast majority being spurious.

FindPOTATOs retrieves real tracklets and is robust against noise. Figure 4 shows the performance per frame triplet (NEAT used a three-image sky survey cadence). from a single night of NEAT TRICAM observations. These observations were taken on 2001 November 24. This night contained 1444 frame triplets of the sky and 1,051,365 total detections. Diagnostic images were generated, but Find_Orb was not used, and there was no stationary source removal by FindPOTATOs. It took 4 hr 37 minutes to process on the Massachusetts Green High Performance Computing Cluster. Each frame triplet had an average of 728 sources, though one frame had 13,319 sources. FindPOTATOs returned 3218 tracklets, 3006 which were determined to be legitimate after visual screening of the diagnostic images and were submitted to the MPC. When provided all needed detections to form a tracklet, FindPOTATOs found the tracklets originally reported by NEAT to the MPC in 2001. Some NEAT-reported tracklets were not found by FindPOTATOs; in all those cases, a detection needed for the tracklet was not provided to FindPOTATOs as input.[9] It additionally identified 1374 legitimate tracklets that were not reported by the NEAT team in 2001, which we also reported to the MPC.

FindPOTATOs was tested against other data sets. We assembled a sample set of observations from the Catalina Sky Survey (A. J. Drake et al. 2009) and ATLAS (J. L. Tonry et al. 2018) from the MPC's database. We also generated synthetic observations of a set of trans-Neptunian objects using JPL Horizons, following a Pan-STARRS-esque observing cadence. With the proper parameters FindPOTATOs can retrieve the tracklets in those samples, including the close approaching NEOs of ATLAS. The fast-moving ATLAS targets require broader parameters compared to NEAT or CSS data. This includes a broader variance in magnitude (4 compared with 2), a faster max_speed ($0\rlap{.}''1$ compared to $0\rlap{.}''05$), a higher velocity_metric_threshold (1–0.25) and a lower min_tracklet_angle (120°–160°). Although these data sets do not include noise, they serve as proof of concept that the code can be applied to varying observing cadences and object orbit types. These test data sets, as well as a sample triplet of reprocessed NEAT data with noise, are available to users in the FindPOTATOs GitHub repository.

## 5. Use of FindPOTATOs

We provide users with example data sets in the GitHub repository, so that they can familiarize themselves with the code's performance on a variety of data. FindPOTATOs is most effective when parameters have been carefully chosen for the specific data set. It is therefore important to use a test set of data with known tracklets to ensure that the chosen parameters return the desired tracklets.

Users may wish to create a custom test set of data before applying FindPOTATOs to their full data set. Test sets can be assembled from a subset of data that contains known detections, by using archived observations from the Minor Planet Center, or by generating synthetic observations using an ephemerides service such as JPL Horizons.

Each data set has a particular blend of desired data and noise. It is important that the test data used in parameter selection has noise that resembles the full data set. Parameter selection generally represents a trade-off between reducing the number of spurious tracklets to a reasonable amount (whatever that is for the user) while preserving the targets of interest. For example, as users can see in the provided example sets, a max_mag_variance of 2 preserved all tracklets from the CSS sample. In that case, that narrow limit may be useful in reducing spurious tracklets for that data set. But that same limit of 2 mag eliminates real tracklets from the close-approaching ATLAS set.

If the max_speed parameter is too high, especially in the case of high source density data, FindPOTATOs will start linking every detection to every other detection. This negatively impacts performance and quality of tracklets. If this happens, max_speed should be reduced.

Some minor planets move roughly linearly across the sky over short timescales. If only sources with linear movement are desired, the min_tracklet_angle parameter can be set to a value such as 170°, meaning that the angle between any three sources in the tracklet must be more than that value. For some use cases, such as some searches for trans-Neptunian objects, objects will appear to reverse course across the plane of the sky due to the relative motion of the object and the observer. To find those objects and not be restricted to quasi-linear sky motion, min_tracklet_angle can be set to 0.

As with all data analyses, it is important to start with a data set that is as clean as possible. This includes stationary star removal and the removal of other forms of noise, such as glints. The SDSS sky image is provided as a tool to help ensure stationary sources are not present in the tracklet; however, it is the user's responsibility to ensure data quality.

While gaining familiarity with the code on a new data set, it is recommended that users check their tracklets against the MPC's Minor Planet Checker (see footnote 7) to confirm that known objects are being found. Confirmation of tracklets of known objects in the data set should be established before tracklets of unknown objects are submitted to the MPC.

---

[9] Some NEAT MPC-reported detections align with known stars and are excluded before FindPOTATOs processing.

# 6. Conclusion

FindPOTATOs is a flexible and robust linking algorithm. It has been shown to find minor planet tracklets in noisy data. It produces tracklets in the two approved MPC output formats, ADES and 80-character, and produces diagnostic tracklet images to assist in quality assurance of tracklets before submission to the MPC. FindPOTATOs is written in Python, to allow users to adapt the code to their data set and add custom features if needed. FindPOTATOs is being applied to the reprocessing of NEAT data, retrieving known tracklets when provided with all needed detections and producing new tracklets not previously reported from that data set.

## Acknowledgments

Support for this work was provided by NASA grant 80NSSC19K1585. The authors are pleased to acknowledge that the computational work reported on in this paper was performed on the shared computing cluster which is administered by New England Research Cloud (NERC, https://nerc.mghpcc.org). We are deeply grateful to the MGHPCC. We thank Bill Gray, creator of Find_Orb, for creating and publishing that software, which is used in this work. We are grateful to the reviewers, whose insightful comments improved this manuscript. We also wish to thank the Minor Planet Center and their devotion to data quality.

## ORCID iDs

C. R. Nugent  https://orcid.org/0000-0003-2504-7887
Nicole J. Tan  https://orcid.org/0000-0001-6541-8887
James M. Bauer  https://orcid.org/0000-0001-9542-0953